\documentclass[preprint,prd,aps,showpacs,showkeys,nofootinbib]{revtex4-1}
\usepackage{amssymb}
\usepackage{amsmath}
\usepackage{graphicx}
\usepackage{subfigure}
\usepackage{float}
\usepackage{mathrsfs}
\usepackage{amsmath}
\usepackage{epstopdf}
\newcommand{\be}{\begin{eqnarray}}
\newcommand{\ee}{\end{eqnarray}}
\newcommand{\PP}{{\cal P}_{\cal R}}

\begin{document}


\title{Gravitational Waves from Double-inflection-point Inflation}

\author{Wu-Tao Xu$^{1,2}$}
\email{xwutao@itp.ac.cn}

\author{Jing Liu$^{1,2}$}
\email{liujing@itp.ac.cn}

\author{Tie-Jun Gao$^3$}
\email{tjgao@xidian.edu.cn}

\author{Zong-Kuan Guo$^{1,2}$}
\email{guozk@itp.ac.cn}

\affiliation{$^1$CAS Key Laboratory of Theoretical Physics, Institute of Theoretical Physics,
 Chinese Academy of Sciences, P.O. Box 2735, Beijing 100190, China}
\affiliation{$^2$School of Physical Sciences, University of Chinese Academy of Sciences,
 No.19A Yuquan Road, Beijing 100049, China}
\affiliation{$^3$School of Physics and Optoelectronic Engineering, Xidian University, Xi'an 710071, China}

\begin{abstract}
We study the production of gravitational waves from primordial scalar perturbations in double-inflection-point inflation,
in which one of the inflection points predicts the power spectra consistent with CMB observations at large scales
and the other generates a large peak in the power spectrum of scalar perturbations at small scales.
We calculate the energy spectrum of the reduced gravitational waves
and find that the gravitational-wave signal can be detected by future space-based laser interferometers.
\end{abstract}

\maketitle

\section{Introduction \label{sec:intr}}

The important prediction of inflation is the generation of primordial scalar and tensor perturbations,
which provide a natural way to explain the anisotropies of the cosmic microwave background (CMB) radiation
and the initial tiny seeds of the large-scale structure.
The existence of primordial scalar perturbations at large scales has firmly been established
by measurements of CMB anisotropies~\cite{Akrami:2018odb}.
Although tensor perturbations at large scales have not been detected,
the Planck 2018 data in combination with BICEP2/Keck Array give a tighter upper limit on the tensor-to-scalar ratio,
$r < 0.064$, at 95\% confidence level~\cite{Akrami:2018odb}.
Detection of such tensor perturbations is regarded as a smoking-gun signature of inflation.

At first order in perturbation theory, the equation of motion for tensor perturbations is a free wave equation.
It means their evolution is not directly influenced by the energy content of the Universe but the cosmic background evolution.
However, at second order the free wave equation gets a source term of first-order scalar perturbations~\cite{Matarrese:1997ay,Acquaviva:2002ud}.
Scalar perturbations that enter the Hubble radius in the radiation-dominated era
can lead to the production of second-order gravitational waves (GWs).
Assuming a power-law spectrum of scalar perturbations at all scales,
such a second-order contribution is generically negligible compared to first-order GWs produced during inflation~\cite{Ananda:2006af,Baumann:2007zm}
because the spectral index of scalar perturbations is constrained at large scales by the Planck 2018 data,
$n_s=0.9649\pm0.0042$ at 68\% confidence level~\cite{Akrami:2018odb}.
However, if the power spectrum of scalar perturbations is enhanced at small scales,
the second-order contribution plays a non-negligible role~\cite{Assadullahi:2009nf,Alabidi:2012ex,Alabidi:2013wtp,Kohri:2018awv,Cai:2018dig,Inomata:2018epa,Cai:2019amo}.
Moreover, the enhancement of scalar perturbations at small scales leads to the production of primordial black holes
via gravitational collapse in the radiation-dominated era~\cite{Yokoyama:1995ex,GarciaBellido:1996qt,Clesse:2015wea,Garcia-Bellido:2016dkw,Garcia-Bellido:2017mdw,Gong:2017qlj,Dalianis:2018frf,Gao:2018pvq}.
Therefore, the induced GWs are used to constrain on the abundance of primordial black holes and vice versa~\cite{Saito:2008jc,Bugaev:2009zh,Saito:2009jt,Bugaev:2010bb,Inomata:2016rbd,Orlofsky:2016vbd}.

The enhancement of the power spectrum of scalar perturbations at small scales can be realized
in a single-field inflationary model with an inflection point~\cite{Garcia-Bellido:2017mdw}.
Actually the inflection-point inflationary model is constructed in supergravity
by using a logarithmic K$\ddot{\rm a}$hler potential and cubic superpotential~\cite{Gao:2015yha}.
Recently a double-inflection-point inflationary model is proposed in supergravity
with a single chiral superfield~\cite{Gao:2018pvq}.
In the model with a double-inflection-point potential,
one of the inflection points can predict the power spectra consistent with the CMB constraints at large scales.
The other inflection point can generate a large peak in the power spectrum of scalar perturbations at small scales,
which leads to the production of primordial black holes in the radiation-dominated era.
In this paper, we shall investigate the production of GWs induced by primordial scalar perturbations
with a large peak at small scale and calculate the GW energy spectrum
in the double-inflection-point inflationary model.

The paper is organized as follows.
In the next section, we briefly review the setup of the double-inflection-point inflationary model in supergravity
and derive primordial power spectrum of scalar perturbations.
In Sec.~\ref{sec:gw}, we present the formalism of the GWs induced by first-order scalar perturbations
and then apply it to calculate the energy spectrum of the reduced GWs in double-inflection-point inflation.
The last section is devoted to summary.

\section{Double-inflection-point inflation \label{sec:infl}}

In this section, we shall setup the double-inflection-point inflationary model in supergravity
and calculate numerically primordial power spectrum of scalar perturbations.
Consider a  K\"{a}hler potential with a shift symmetry~\cite{Ketov:2016gej}
\begin{eqnarray}
&&K=ic (\Phi-\bar{\Phi})-\frac{1}{2}(\Phi-\bar{\Phi})^2-\frac{\zeta}{4}(\Phi-\bar{\Phi})^4,
\label{kp6}
\end{eqnarray}
where $c$ and $\zeta$ are two real parameters. The real component $\phi$ of the chiral superfield $\Phi=(\phi+i\chi)/\sqrt{2}$ is taken to be the
inflaton and the quartic term $\frac{\zeta}{4}(\Phi-\bar{\Phi})^4$ serves to stabilize the field $\phi$ during inflation at $\langle\chi\rangle \simeq 0$ by making $\zeta$ sufficient large.

Let us consider an exponential superpotential of the form
\begin{eqnarray}
&&W=a_0(1+a_1 e^{-b_1 \Phi }+a_2 e^{-b_2 \Phi }+a_3 e^{-b_3 \Phi }).
\label{infp}
\end{eqnarray}
Such a kind of superpotentials with exponential functions with two terms have been studied
in the so-called racetrack model~\cite{Krasnikov:1987jj,Escoda:2003fa,BlancoPillado:2004ns}
and in other models~\cite{Ketov:2016gej}.
If we restore the SUSY in vacuum with a vanishing cosmological constant
(The issue of SUSY breaking in vacuum is discussed in Ref.~\cite{Gao:2015yha}),
the F-term should be vanished $D_{\Phi}W=0$, and $V=0$ at $\Phi=0$, which requires the constraint
\begin{eqnarray}
&&W=\partial_{\Phi}W=0.
\end{eqnarray}
Then two of the parameters, $a_1$ and $a_2$, can be eliminated by solving the constraint as
\begin{eqnarray}
&&a_1\to \frac{b_2+a_3 b_2-a_3 b_3}{b_1-b_2},\quad a_2\to \frac{-b_1-a_3 b_1+a_3 b_3}{b_1-b_2}.
\end{eqnarray}

Substituting the  K\"{a}hler potential and superpotential  into
\begin{eqnarray}
&&V=e^{K/M_P^2}\Big[D_{\Phi_i}W(K^{-1})^{ij^*}D_{\Phi_j^*}W^*-3M_P^{-2}|W|^2\Big],
\label{infp}
\end{eqnarray}
where
\begin{eqnarray}
&&D_{\Phi}W=\partial_{\Phi}W+M_P^{-2}(\partial_{\Phi}K)W,
\end{eqnarray}
and $(K^{-1})^{ij^*}$ is the inverse of the K\"{a}hler metric
\begin{eqnarray}
&&K^{ij^*}=\frac{\partial^2K}{\partial\Phi_i\partial\Phi^*_{j}},
\end{eqnarray}
one can get the scalar potential $V(\phi)$.

\begin{figure}
\includegraphics[width=4in]{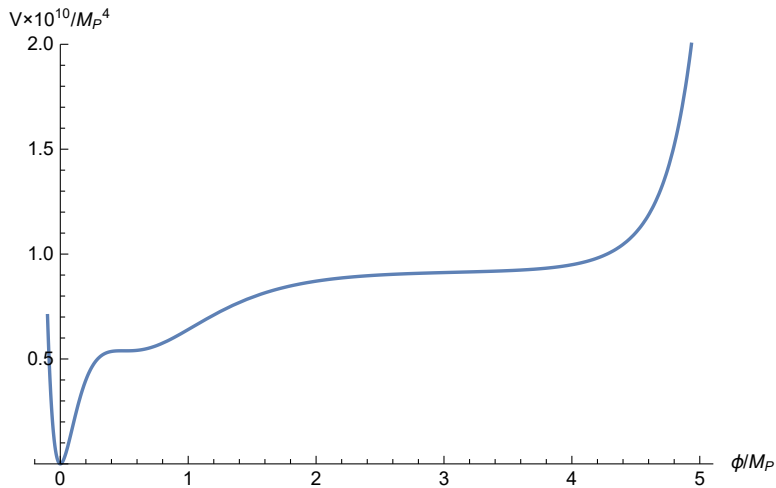}
\caption{Scalar potential $V(\phi)$ with double inflection points.}
\label{fig:potential}
\end{figure}

In some choices of parameter space, for instance we take
\begin{eqnarray}
 a_0=4.35\times10^{-6}, a_3=7\times10^{-8}, b_1=3.05, b_2=6.3868164, b_3=-4.4, c=2.8.
\label{eq:para}
\end{eqnarray}
Such a potential $V(\phi)$ have two nearly inflection points, as shown in Fig.~\ref{fig:potential}.
One of the inflection points at large scales can make the prediction of the scalar spectral index and tensor-to-scalar ratio
consistent with the current CMB data,
and the other one at small scales can generate a large peak in the power spectrum of scalar perturbations to arise primordial black holes~\cite{Gao:2018pvq}.

In the single-field slow-roll framework of a FRW background,
the slow-roll parameters $\epsilon_V$ and $\eta_V$ can be calculated as
\begin{eqnarray}
&&\epsilon_V=\frac{M_P^2}{2}\Big(\frac{V'}{V}\Big)^2,\nonumber\\
&& \eta_V =M_P^2\Big(\frac{V''}{V}\Big).
\end{eqnarray}
However, near the inflection point, the potential becomes extremely flat,
thus the slow-roll approximation is no longer applicable~\cite{Germani:2017bcs,Dimopoulos:2017ged},
and the ultra-slow-roll trajectory supersedes the slow-roll one.\footnote{It has been point out that
quantum diffusion can  enhance the power spectrum near the inflection point~\cite{Ezquiaga:2018}, however, this
was questioned in ~\cite{Cruces:2018}. So details of the quantum diffusion effect on PBH formation are still
under debate.}
Thus one has to use the slow-roll parameters defined by the Hubble parameter instead~\cite{Schwarz:2001vv,Leach:2002ar,Schwarz:2004tz},
\begin{eqnarray}
&&\epsilon_H=-\frac{\dot{H}}{H^2},\nonumber\\
&&\eta_H=-\frac{\ddot{H}}{2H\dot{H}}=\epsilon_H-\frac{1}{2}\frac{d\ln\epsilon_H}{dN_e},\nonumber\\
&&\xi_H=\frac{\dddot{H}}{2H^2\dot{H}}-2\eta_H^2=\epsilon_H \eta_H-\frac{d\eta_H}{dN_e},
\end{eqnarray}
where  primes denote derivatives with respect to the field $\phi$ and dots represent derivatives with respect to cosmic time,
$N_e(t)$ is the $e$-folding number
between the crossing time of the $k_{*}$ scale and the time of the inflation end, which is required  in the range of $50-60$.

The scalar spectral index and its running as well as the tensor-to-scalar ratio can be expressed at the leading order using $\epsilon_H$, $\eta_H$ and $\xi_H$  as
\begin{eqnarray}
&&n_s=1-4\epsilon_H+2\eta_H,\nonumber\\
&&\alpha=\frac{dn_s}{d\ln k}=10\epsilon_H \eta_H-8 \epsilon_H^2-2 \xi_H, \nonumber\\
&&r=16\epsilon_H.
\end{eqnarray}
For the parameter set~\eqref{eq:para}, the numerical results are
\begin{eqnarray}
&n_s=0.9635,\quad \alpha=-0.00369,\quad r=0.00276,
\end{eqnarray}
which are in agreement with the  current  CMB constraints  from Planck 2018 at $68\%$ confidence level
at the pivot scale $k_*=0.05\mathrm{Mpc}^{-1}$~\cite{Akrami:2018odb}
\begin{eqnarray}
&&n_s=0.9640\pm0.0043,\quad  \alpha=-0.0071\pm0.0068,\quad r<0.079.
\end{eqnarray}

\begin{figure}
\includegraphics[width=4in]{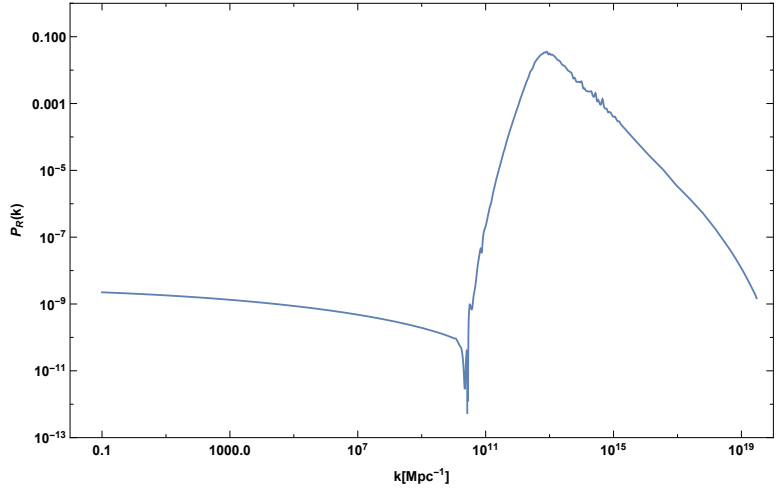}
\caption{Primordial power spectrum of scalar perturbations $\PP (k)$
predicted by the double-inflection-point inflationary model.}
\label{fig:pert}
\end{figure}

However,  for the power spectrum of scalar perturbations,
it is pointed out that near the inflection point calculation using the approximate expression $\PP \simeq\frac{1}{8\pi^2M_P^2}\frac{H^2}{\epsilon_H}$ will underestimates the power spectrum~\cite{Gao:2018pvq}.
Thus one must solve the MS equation of mode function numerically,
\begin{eqnarray}
&&\frac{d^2u_k}{d\eta^2}+\Big(k^2-\frac{1}{z}\frac{d^2z}{d\eta^2}\Big)u_k=0,
\end{eqnarray}
where $\eta$ is the conformal time and $z\equiv\frac{a}{\mathcal{H}}\frac{d\phi}{d\eta}$.
And the initial condition is taken to be the Bunch-Davies type~\cite{Bunch:1978yq}
\begin{eqnarray}
&&u_k\rightarrow \frac{e^{-ik\eta}}{\sqrt{2k}},\;\;  \text{as}\;\; \frac{k}{aH}\rightarrow\infty.
\end{eqnarray}
For the purpose of numerical simulation, the MS equation can be written in terms of $N_e$ as the time variable~\cite{Ballesteros:2017fsr}
\begin{eqnarray}
&&\frac{d^2u_k}{dN_e^2}+(1-\epsilon_H)\frac{du_k}{dN_e}+[\frac{k^2}{\mathcal{H}^2}+(1+\epsilon_H-\eta_H)(\eta_H-2)-\frac{d\epsilon_H-\eta_H}{dN_e}]=0,
\label{eq:ms}
\end{eqnarray}
and the primordial power spectrum can be calculated by
\begin{eqnarray}
&&\PP=\frac{k^3}{2\pi^2}\Big|\frac{u_k}{z}\Big|^2_{k\ll\mathcal{H}}.
\end{eqnarray}
The numerical results for the parameter set~\eqref{eq:para} calculated from the solutions to the MS Eq.~\eqref{eq:ms} is plotted in Fig.~\ref{fig:pert}.
We can see that there is a large peak at small scales in the spectrum,
with a height of about seven orders of magnitude more than the spectrum at CMB scales,
which can generate primordial black holes via gravitational collapse.

\section{Induced gravitational waves \label{sec:gw}}

In this section, we first present the formalism of the GWs induced by first-order scalar perturbations
and then give numerical results by using the primordial power spectrum obtained in Sec.~\ref{sec:infl}.
The predicted energy spectrum of GWs exceeds the sensitivity curve for LISA.
This implies that the GW signal can be detected by LISA.
Furthermore, we generalize this result to the primordial power spectrums with a power-law form.
\subsection{Basic Equations}

In the conformal Newtonian gauge the metric can be written as
\begin{equation}
\label{eq:metric}
d s^{2}=a^{2}(\eta)\left\{-(1+2 \Psi) d \eta^{2}+\left[(1-2 \Psi) \delta_{i j}+\frac{1}{2} h_{i j}\right] d x^{i} d x^{j}\right\},
\end{equation}
where $\Psi$ is the first-order scalar perturbations and $h_{ij}$ are tensor perturbations.
Here, we have neglected the first-order GWs, vector perturbations and anisotropic stress~\cite{Baumann:2007zm}.

The Fourier modes of tensor perturbations are introduced as usual
\begin{equation}
\label{eq:fourier}
h _ { i j } (\eta, \mathbf{x}) = \int \frac { d^3 \mathbf{k} } { ( 2 \pi ) ^ { 3 / 2 } } e ^ { i \mathbf{k} \cdot \mathbf{x} } \left[ h_\mathbf{k}^ { + } ( \eta ) \mathrm{e}_{ij}^+ ( \mathbf{k} ) + h_\mathbf{k}^\times ( \eta ) \mathrm{e}_{ij}^\times (\mathbf{k}) \right],
\end{equation}
where $\mathrm{e}_{ij}^+ (\mathbf{k})$ and $\mathrm{e}_{ij}^\times (\mathbf{k})$ are polarization tensors that satisfy $\sum _ { i,j } \mathrm { e } _ {ij}^\alpha (\mathbf{k}) \mathrm{e}_{ij}^\beta (-\mathbf{k}) =  \delta ^ { \alpha \beta }$.
We shall omit the polarization index in the following.

The equation of motion of tensor modes in the momentum space can be derived from the Einstein equation to second order
\begin{equation}
\label{eq:tensoreom}
h_{\mathbf{k}}^{\prime \prime}+2 \mathcal{H} h_{\mathbf{k}}^{\prime}+k^{2} h_{\mathbf{k}}=S(\eta,\mathbf{k}),
\end{equation}
where $S(\eta, \mathbf{k})$ is the Fourier transformation of the source term $S_{i j}(\eta, \mathbf{x})$,
 \begin{equation}
\label{eq:tensoreomf}
S(\eta, \mathbf{k}) = -4 {\mathrm e}^{i j}(\mathbf{k}) \int \frac{d^3 \mathbf{x}}{(2 \pi)^{3 / 2}} e^{-i \mathbf{k} \cdot \mathbf{x}} S_{i j}(\eta, \mathbf{x}).
\end{equation}
The source term is given by~\cite{Baumann:2007zm,Ananda:2006af}
\begin{equation}
\label{eq:source}
S_{i j}(\eta, \mathbf{x}) = 4 \Psi \partial_{i} \partial_{j} \Psi+2 \partial_{i} \Psi \partial_{j} \Psi-\frac{4}{3(1+w) \mathcal{H}^{2}} \partial_{i}\left(\Psi^{\prime}+\mathcal{H} \Psi\right) \partial_{j}\left(\Psi^{\prime}+\mathcal{H} \Psi\right).
\end{equation}
The equation of motion can be formally solved using the Green's function method and the solution reads
\begin{equation}
\label{eq:greensolu}
h_{\mathbf{k}}(\eta)=\frac{1}{a(\eta)} \int^{\eta} d \tilde{\eta} G_k(\eta ; \tilde{\eta})[a(\tilde{\eta}) S(\tilde{\eta}, \mathbf{k})],
\end{equation}
where $G_k(\eta ; \tilde{\eta})$ is the solution of
\begin{equation}
\label{eq:deltafunc}
\dfrac{d^2 G_k (\eta ; \tilde{\eta})}{d \tilde{\eta}^{2}}+\left(k^{2}-\frac{d^2 a}{a d\tilde{\eta}^{2}}\right) G_k (\eta ; \tilde{\eta})=\delta(\eta-\tilde{\eta}).
\end{equation}
To calculate the evolution of the source term, we need to solve the equation of motion of scalar perturbations~\cite{Kodama:1985bj,Mukhanov:1990me}
\begin{equation}
\label{eq:scalareom}
\Psi_{\mathbf{k}}^{\prime \prime}(\eta)+\frac{6(1+w)}{(1+3 w) \eta} \Psi_{\mathbf{k}}^{\prime}(\eta)+w k^{2} \Psi_{\mathbf{k}}(\eta)=0,
\end{equation}
where $w$ is the equation of state of the Universe. In the following, we split $\Psi_{\mathbf{k}}(\eta)$ into the primordial value $\psi_{\mathbf{k}}$ and the transfer function $\Psi(k\eta)$
\begin{equation}
\label{eq:split}
\Psi_{\mathbf{k}}(\eta) \equiv\Psi(k \eta) \psi_{\mathbf{k}}.
\end{equation}
Then the Fourier transform of the source term can be written as
\begin{equation}
\label{eq:sourcedet}
S(\eta, \mathbf{k})=\int \frac{d^3 \mathbf{p}}{(2 \pi)^{3 / 2}} \mathrm{e} (\mathbf{k}, \mathbf{p}) f(\eta, \mathbf{k}, \mathbf{p}) \psi_{\mathbf{k}} \psi_{\mathbf{k}-\mathbf{p}},
\end{equation}
where
\begin{equation}
\label{eq:eij}
\mathrm{e}(\mathbf{k}, \mathbf{p}) = \mathrm{e}^{i j}(\mathbf{k}) p_{i} p_{j},
\end{equation}
and
\begin{equation}
\label{eq:fkq}
\begin{split}
 f(\eta, \mathbf{k}, \mathbf{p}) = &\frac{8(3 w+5)}{3(w+1)} \Psi(|\mathbf{p}| \eta) \Psi(|\mathbf{k}-\mathbf{p}| \eta)+\frac{4(3 w+1)^{2}}{3(w+1)} \eta^{2} \Psi^{\prime}(|\mathbf{p}| \eta) \Psi^{\prime}(|\mathbf{k}-\mathbf{p}| \eta) \\
 &+\frac{8(3 w+1)}{3(w+1)} \eta\left[\Psi^{\prime}(|\mathbf{p}| \eta) \Psi(|\mathbf{k}-\mathbf{p}| \eta)+\Psi(|\mathbf{p} | \eta) \Psi^{\prime}(|\mathbf{k}-\mathbf{p}| \eta)\right].
 \end{split}
\end{equation}
For the modes well inside the horizon, the energy spectrum of GWs can be expressed in terms of the power spectrum of GWs
\begin{equation}
\label{eq:OmegaGW}
\Omega_{\mathrm{GW}}(\eta, k)\equiv\dfrac{1}{\rho_{c}}\dfrac{d\rho_{\mathrm{GW}}}{d\ln k} = \frac{1}{24}\left(\frac{k}{\mathcal{H}}\right)^{2} \overline{{\mathcal P}_{h}(\eta, k)}\,,
\end{equation}
where the overscore denotes the oscillation average and the two polarization modes are summed up
and $\rho_{c}$ is the critical energy density of the Universe.
The power spectrum of GWs, ${\mathcal P}_{h}$, is defined as
\begin{equation}
\label{eq:Ph}
\left\langle h_{\mathbf{k}}(\eta) h_{\mathbf{p}}(\eta)\right\rangle= \frac{2 \pi^{2}}{k^{3}} \delta^{3}(\mathbf{k}+\mathbf{p}) {\mathcal P}_{h}(\eta, k).
\end{equation}
To calculate the two-point correlation function of $h_{\mathbf{k}}$, using Eq.~\eqref{eq:greensolu}
and Eq.~\eqref{eq:sourcedet} one can obtain
\begin{equation}
\label{eq:twopoint}
\left\langle h_{\mathbf{k}}(\eta) h_{\mathbf{p}}(\eta)\right\rangle =
\int \frac{d^3 q \; d^3 \tilde{q}}{(2 \pi)^3} \mathrm{e}(\mathbf{k}, \mathbf{q})  \mathrm{e}(\mathbf{p}, \tilde{\mathbf{q}}) I(\eta,\mathbf{k}, \mathbf{q}) I(\eta, \mathbf{p}, \tilde{\mathbf{q}})\left\langle \psi_{\mathbf{q}} \psi_{\mathbf{k}-\mathbf{q}} \psi_{\tilde{\mathbf{q}}} \psi_{\mathbf{p}-\tilde{\mathbf{q}}}\right\rangle,
\end{equation}
where
\begin{equation}
\label{eq:I}
I(\eta, \mathbf{k}, \mathbf{p}) \equiv \int^{\eta} d \tilde{\eta} \frac{a(\tilde{\eta})}{a(\eta)} G_k (\eta ; \tilde{\eta}) f(\tilde{\eta}, \mathbf{k}, \mathbf{p}).
\end{equation}
The four-point correlation function of $\psi_{\mathbf{k}}$ can be transformed into the two-point correlation function
once $\psi_{\mathbf{k}}$ is assumed to be Gaussian according to Wick's theorem.
Utilizing the dimensionless variables $u \equiv|\mathbf{k}-\mathbf{p}| / k$, $v \equiv|\mathbf{p}| / k$ and  $x \equiv k \eta$, finally we obtain~\cite{Kohri:2018awv}
\begin{equation}
\label{eq:twopoint2}
{\mathcal P}_{h}(\eta, k)=4 \int_{0}^{\infty} d v \int_{|1-v|}^{1+v} d u\left(\frac{4 v^{2}-\left(1+v^{2}-u^{2}\right)^{2}}{4 u v}\right)^{2} \mathcal{I}^{2}(x, u, v) \PP (k u) \PP (k v),
\end{equation}
where
\begin{equation}
\label{eq:I2}
\mathcal{I}(x, u, v) \equiv I(\eta, \mathbf{k}, \mathbf{p}) k^{2}.
\end{equation}
For the power spectrum of scalar perturbations in double-inflection-point inflation obtained in Sec.~\ref{sec:infl},
the peak mode enters the horizon in the radiation-dominated era, i.e., $\omega=1/3$. The solution to Eq.~\eqref{eq:scalareom} is simplified to
\begin{equation}
\label{eq:solutontos}
\Psi(x)=\frac{9}{x^{2}}\left(\frac{\sin (x / \sqrt{3})}{x / \sqrt{3}}-\cos (x / \sqrt{3})\right),
\end{equation}
and the Green's function is
\begin{equation}
\label{eq:greens}
G_k (\eta; \tilde{\eta})=\dfrac{\sin (k\eta-k\tilde{\eta})}{k}.
\end{equation}
The peak corresponds to the scale of $10^{13}$ Mpc$^{-1}$,
many orders of magnitude larger than the pivot scale $k_{*}=0.05\mathrm{Mpc}^{-1}$~\cite{Akrami:2018odb}.
Since the scalar perturbations are damped at sub-horizon scales in the radiation-dominated era as shown in Eq.~\eqref{eq:solutontos},
the production of the induced GWs mostly occur around the time when the corresponding mode crosses the horizon.
It allows us to take advantage of the method applied in~\cite{Kohri:2018awv}.
One can obtain the oscillation average taking the late-time limit $x \rightarrow \infty$
\be
&& \overline{\mathcal{I}^{2}(x \rightarrow \infty, u, v)}
 = \dfrac{1}{2}\left(\frac{3}{4 u^{3} v^{3} x}\right)^{2}\left(u^{2}+v^{2}-3\right)^{2} \nonumber \\
&& \quad \left\{ \left[-4 u v+\left(u^{2}+v^{2}-3\right) \ln \left|\frac{3-(u+v)^{2}}{3-(u-v)^{2}}\right|\right]^{2}
 +\left[\pi\left(u^{2}+v^{2}-3\right) \Theta(u+v-\sqrt{3})\right]^{2} \right\},
\label{eq:I3}
\ee
where $\Theta$ is the Heaviside theta function. Together with Eq.~\eqref{eq:Ph} and Eq.~\eqref{eq:I2},
using $\mathcal{H}=1/\eta$ in the radiation-dominated era, we finally obtain
\be
&& \Omega_{\mathrm{GW}}(\eta,k)
 = \dfrac{1}{12}\int_{0}^{\infty} dv \int_{|1-v|}^{1+v} du \left(\frac{4 v^{2}-\left(1+v^{2}-u^{2}\right)^{2}}{4 u v}\right)^{2} \PP (k u) \PP (k v) \nonumber \\
&& \quad \left(\frac{3}{4 u^{3} v^{3} }\right)^{2}\left(u^{2}+v^{2}-3\right)^{2} \nonumber \\
&& \quad \left\{\left[-4 u v+\left(u^{2}+v^{2}-3\right) \ln \left|\frac{3-(u+v)^{2}}{3-(u-v)^{2}}\right|\right]^{2}+\left[\pi\left(u^{2}+v^{2}-3\right) \Theta(u+v-\sqrt{3})\right]^{2} \right\}.
\label{eq:final}
\ee

\subsection{Numerical Results and Observational Implications}
\begin{figure}
\includegraphics[width=4in]{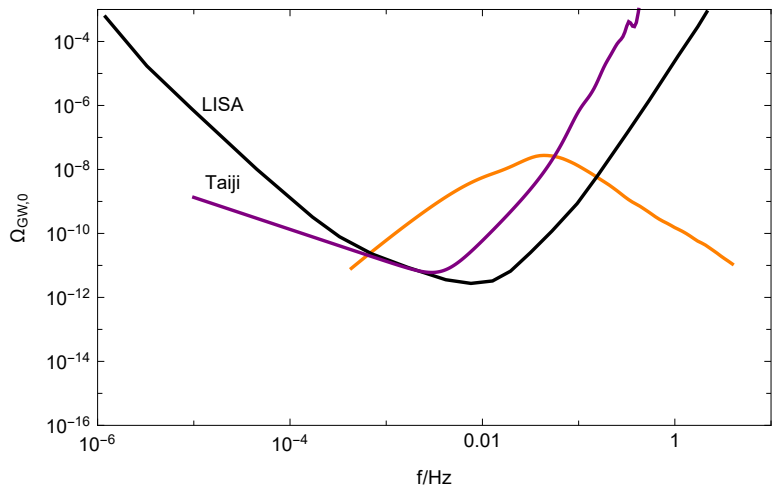}
\caption{Energy spectrum of the induced gravitational waves $\Omega_{\rm GW,0}$
predicted by the double-inflection-point inflationary model (the orange curve).
The black/red curves are the expected sensitivity curve of LISA~\cite{Audley:2017drz} and Taiji~\cite{Guo:2018npi}, respectively.}
\label{fig:gwLisa}
\end{figure}

In this subsection, we shall obtain the numerical results of the induced GWs in the double-inflection-point inflationary model
and compare the results of $\Omega_{\mathrm{GW}}$ to the expected sensitivity curves of LISA~\cite{Audley:2017drz}
and Taiji~\cite{Guo:2018npi}.
The energy spectrum at the present time $\Omega_{\mathrm{GW}, 0}$ is related to one produced in the radiation-dominated era as
\begin{equation}
\label{eq:Omegaappr}
\Omega_{\mathrm{GW}, 0}=\Omega_{r, 0}\left(\frac{g_{*,0}}{g_{*,p}}\right)^{1 / 3} \Omega_{\mathrm{GW}},
\end{equation}
where $\Omega_{r, 0}$ is the density fraction of radiation today,
$g_{*,0}$ and $g_{*,p}$ are the the effective numbers of relativistic degrees of freedom at the present time
and at the time when the peak mode crosses the Hubble horizon, respectively.
The present value of the frequency $f$ is
\begin{equation}
\label{eq:Omegaappr}
f \approx 0.03 \mathrm{Hz} \frac{k}{2\times 10^7 \mathrm{pc}^{-1}}\,.
\end{equation}
From Fig.~\ref{fig:gwLisa} we can see that the numerical result lies above the expected sensitivity curves of
LISA~\cite{Audley:2017drz} and Taiji~\cite{Guo:2018npi}.
Compared to LISA, the sensitivity curve of Taiji shifts to low frequency due to larger spacecraft separations.
The predicted peak frequency, about $0.05$ Hz, is within the frequency range of space-based GW detectors.
Moreover, for the modes much smaller or larger than the peak mode $k_{p}$,
$\Omega_{\mathrm{GW}}$ is well described by a power law of $k$.
In the region $k\ll k_p$, $\Omega_{\mathrm{GW}}\propto k^{3}$,
which is the same as other stochastic backgrounds of GWs sourced by preheating and phase transition,
while the index in the region $k\gg k_{p}$ turns out to be $1.9$,
which is distinguished from other GW sources.

\subsection{Results with a power-law power spectrum}
As shown in Fig.~\ref{fig:pert}, the primordial power spectrum can be well approximated by a power-law function of $k$ in the vicinity of the peak.
For $k\ll k_{p}$ and $k\gg k_{p}$, $\PP(k) \propto k^{n_1}$ and $\PP(k) \propto k^{n_2}$, respectively.
In this case, we find $\Omega_\mathrm{GW}(k)\propto k^{m_1}$ and $\Omega_\mathrm{GW}(k)\propto k^{m_2}$.
The power-law exponents $m_1$ and $m_2$ can be expressed in terms of $n_1$ and $n_2$, respectively.
First, we consider the case of $k\ll k_{p}$.
It is expected that the main contribution to $\Omega_{\mathrm{GW}}$ comes from the peak.
Since $v$ and $u$ are inversely proportional to $k$,
the main part of the integral Eq.~\eqref{eq:final} comes from the interval of $1 \ll v,u \le k_p/k$.
In this limit, Eq.~\eqref{eq:final} is simplified to
\begin{equation}
\label{eq:kllkp}
\Omega_{\mathrm{GW}}(k)\propto k^{2n_{1}} \int_{0}^{k_{p}/k}\mathrm{d}v\,v^{2n_{1}-4}.
\end{equation}
If $n_{1}>2$, the integrand is an increasing function of $v$. It is obtained that
\begin{equation}
\label{eq:left}
\Omega_{\mathrm{GW}}(k)\propto k^{3}.
\end{equation}
Second, we consider the case of $k\gg k_{p}$.
The integrand is simplified to $v^{3-n_{2}}$ if $ v\ll1 $ or $ u\ll1 $.
For $n_{2} > -4$, Eq.~\eqref{eq:final} can be expressed as
\begin{equation}
\label{eq:right0}
\Omega_{\mathrm{GW}}(k)\propto k^{2n_{2}} \int_{-\infty}^{+\infty}dv\,F(v)
.
\end{equation}
Since the integration in~\eqref{eq:right0} is independent of $k$, $\Omega_{\mathrm{GW}}(k)\propto k^{2n_{2}}$.
For $n_{2}\le -4$, the interval of $k_p/k \le v\ll1$ or $k_p/k \le u\ll 1$ contributes mainly to the integral.
In this case Eq.~\eqref{eq:final} is then simplified to
\begin{equation}
\label{eq:right}
\Omega_{\mathrm{GW}}(k)\propto k^{2n_{2}} \int_{k_{p}/k}^{\epsilon}\mathrm{d}v\,v^{3-n_{2}},
\end{equation}
where $ \epsilon $ satisfies $k_{p}/k \ll \epsilon \ll 1$.
One has $\Omega_{\mathrm{GW}}(k)\propto k^{n_{2}-4}$ for $n_{2}<-4$ and $\Omega_{\mathrm{GW}}(k)\propto k^{n_{2}-4}\ln(k_{p}/k)$ for $n_{2}=-4$.

\begin{figure*}
\includegraphics[width=3in]{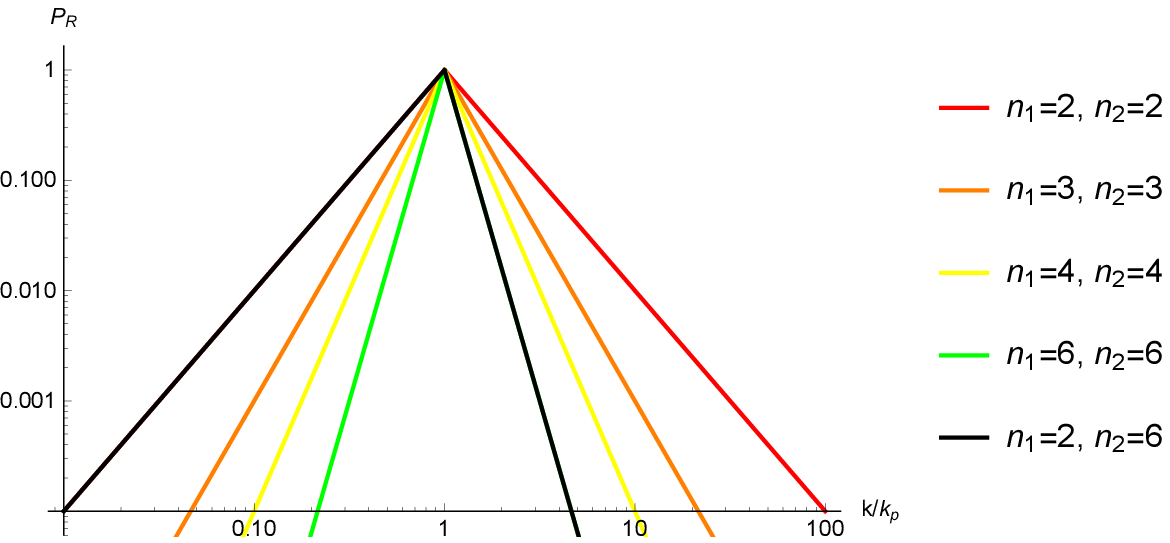}
\includegraphics[width=3in]{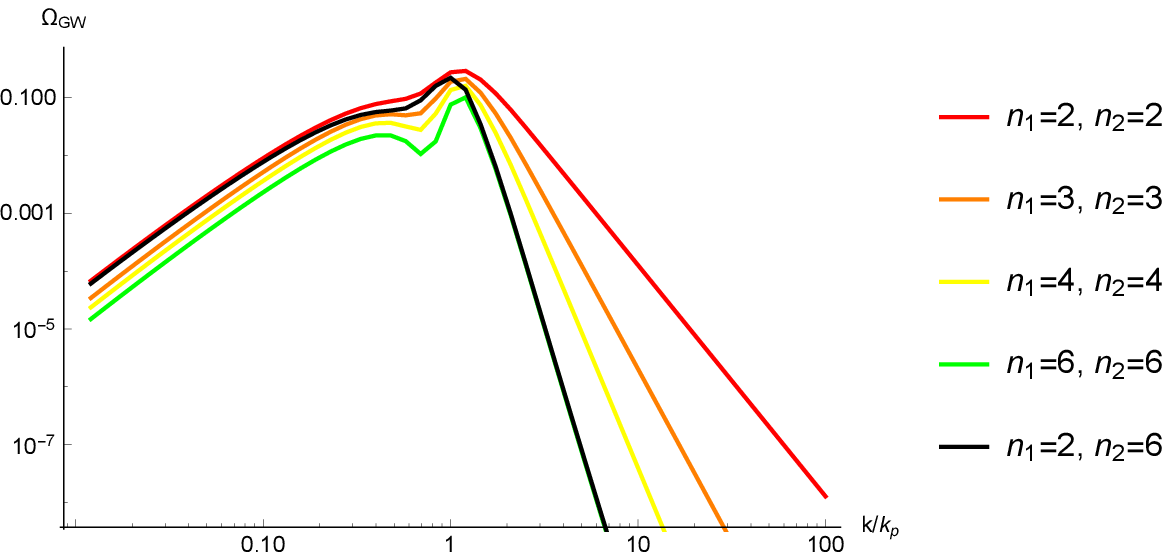}
\caption{Asymptotic power law behavior of $\Omega_{\mathrm{GW}}$ with respect to $k$. In the left panel, we give several examples of power law $P_{\mathcal{R}}$ with different indexes. As depicted in the right panel, the numerical results is in good agreement of the analytic results.}
\label{fig:index}
\end{figure*}

\section{Summary \label{sec:sum}}

In this paper, we have numerically calculated the energy spectrum of the stochastic background of GWs
reduced by first-order scalar perturbations which have a power spectrum with a large peak at small scales
in the double-inflection-point inflationary model.
We have found that the GW signal can be probed by the planned space-based interferometers such as LISA and Taiji.

\begin{acknowledgments}
This work was supported by the National Natural Science Foundation of China Grants No. 11575272, No. 11690021 and No. 11851302.
TJG was supported by ``the National Natural Science Foundation of China'' (NNSFC) with Grant No. 11705133,
and ``the Fundamental Research Funds for the Central Universities'' No.JBF180501.
\end{acknowledgments}

\end{document}